\documentclass[review]{elsarticle}

\usepackage{hyperref}

\journal{ArXiv.org}

\begin{document}

\begin{frontmatter}

\title{BitTorrent is Apt for Geophysical Data Collection and Distribution}

\author[iperasaddress]{K.\,I.~Kholodkov\corref{correspondingauthor}}
\author[iperasaddress]{I.\,M.~Aleshin}
\author[iperasaddress]{S.\,D.~Ivanov}
\address[iperasaddress]{Schmidt Institute of Physics of the Earth of the Russian Academy of Sciences
              123242 Russia, Moscow, Bolshaya Gruzinskaya 10b1}
\cortext[correspondingauthor]{Corresponding author}

\begin{abstract}
This article covers a nouveau idea of how to collect and handle geophysical data with a peer-to-peer network in near real-time. The text covers a brief introduction to the cause, the technology, and the particular case of collecting data from GNSS stations. We describe the proof-of-concept implementation that has been tested. The test was conducted with an experimental GNSS station and a data aggregation facility. In the test, original raw GNSS signal measurements were transferred to the data aggregation center and subsequently to the consumer. Our implementation utilized BitTorrent to communicate and transfer data. The solution could be used to establish the majority of data aggregation centers’ activities to provide fast, reliable, and transparent real-time data handling experience to the scientific community.
\end{abstract}


\end{frontmatter}

\section{Introduction}

Data plays an important role in geophysics. Among researchers, quality and ease of access are its most valued properties~\cite{Borgman2019Lives}. Whether data quality is the merit of the instrument, the site, or the fieldwork, ease of access is often left out. In many cases fieldwork experiments collect and subsequent work consume data without putting sufficient effort into making sure that the source data would be available to the public, for a reasonable amount of time at least.

We believe this predominantly happens because the scientific community relies on legacy general-purpose technology to publish and distribute data. It is worth noting that some scientific communities created and adopted dedicated technology to meet their data acquisition and handling needs. For instance, the seismological research community developed several~\cite{havskov2012seismic} protocols and formats to facilitate real-time reliable data acquisition, storage, and processing. It turned out that some of these protocols and formats could be utilized in adjacent scientific fields of expertise - magnetic observation and structural monitoring.  That was possible because of the nature of the collected data. Each instrument provides one or several time series - a sequence of numeric values bound to a unique timestamp. The protocols are generally agnostic to the meaning of values that are transferred. This way, we previously adopted~\cite{ivanov2019geophysical} the Seedlink protocol to transfer magnetic and structural monitoring data. However, these solutions only cover the immediate data availability issues and leave data sharing and data reuse problems far behind. When the time comes for accessing data years ago the way that historical data is stored comes into play. Among different ways~\cite{saul2014seismic} of getting that data, one can particularly note FTP servers and data ordering with email or automated web form. Both ways feel neither robust nor voguish.

Distributing data across several network sites partially addresses the problem because the data can be found and accessed faster if there is more than one source\cite{adgeo-40-31-2014}. To achieve this, data is replicated among several nodes in the network, and special software is utilized to enable it. The software is mirroring solutions (rsync-like), distributed filesystem or database, or a peer network. In most cases the use of new and complex software is inevitable. Among the solutions that enable distributed data storage and access we particularly highlight peer-to-peer network data-sharing technologies.  These techs have an advantage over others for the task of enabling access to geophysical data - one probably has used them at least once. Indeed, peer networks had great success in delivering data to end users - from updates to operating systems and applications to TV shows. Inherently, we adopted BitTorrent for the needs of geophysical data collection.

\section{Original Problem}

Before adopting BitTorrent we explored other data delivery protocols to find the technology combination that is bandwidth-efficient, tolerant to network failures, and has wide software support. For time-series, we’ve found Seedlink to be the optimal solution~\cite{ivanov2019geophysical}. However, neither Seedlink, nor any other time-series protocols and corresponding file formats can be used efficiently to transfer and store complex data - like GNSS measurements or ionograms, and here’s why. 

Unlike time-series data, these formats contain multiple, in most cases, position-dependent structures. Such structures can be called messages~\cite{greis}, groups~\cite{sao}, packets~\cite{ubx}, etc. Often an instance of a particular structure cannot provide a single complete instrument measurement because it is dependent on either a separate time structure or a similar previous structure, due to differential (or delta) recording. Sometimes, the community offers unified formats, such as RINEX\cite{rinex} to store this data. This kind of unification helps to fight the so-called vendor lock when the user is no longer tied to specific processing software, however, while these solutions combat format diversity they mostly omit transfer issues. With that said we believe it is optimal to conserve the original format for data handling mainly because format unification does not benefit data transfer.

Many data-archiving efforts like IGS~\cite{igs}, INTERMAGNET~\cite{intermagnet}, and weather agencies generally around the globe utilize antiquated FTP and SMTP, the one used for email, to transfer data from stations and among datacenters. These protocols are badly inefficient in terms of bandwidth, accessibility (see FTP ‘active’ mode), and error handling but have vast software support. While FTP shows good transfer speeds in theory, real-life instances are often oversubscribed and limit both speed and number of connections. SMTP adds 30\% on average to the payload, and, even if used with private servers, is prone to delays. Both don’t provide any error detection and correction relying on TCP and the link level, Ethernet or LTE,  to handle this task. However, the undetected error rate is quite observable~\cite{10.1145/347059.347561}. To combat these issues we suggest using peer data-sharing networks. Their protocols are designed to handle large bandwidth, simultaneous delivery, and consistency control on one hand. The drawback, on the other hand, is the lack of production-ready software support for a data acquisition site. In this paper, we describe our experience of adopting a peer network protocol for geophysical data collection and transfer.
\section{The Case}

As part of the effort on establishing a data aggregation center authors had to perform various data collection incentives: tiltmeter installations, seismic stations, magnetic observatories, and GNSS surveys, and so forth. Tiltmeters, seismic stations, and magnetic observatories net time-series data, so we utilize Seedlink and MQTT to establish real-time data acquisition. In contrast, GNSS measurements do not have a well-known guaranteed data transfer solution like Seedlink. From the station’s perspective, the measurements can be stored as bulky files and there should be a way to transfer them reliably to a data center and consumers. The size of files from a single station can reach several gigabytes a day. The size poses a problem for both remote stations’ network access channels and distribution centers. From~\cite{7121446} we also learned that stable connection to the Internet is rather uncommon at remote locations causing data consistency and availability problems.  As shown above, we did not feel comfortable with FTP or SMTP so we decided to find another solution to transfer that amount of data. That solution was to use the BitTorrent peer network.
\section{BitTorrent}

Among several maintained peer network implementations~\cite{Maymounkov}\cite{bt}\cite{Kulbak2005TheEP}\cite{gnu} we have chosen the one with greater software support and overall popularity~\cite{btt}. At a glance, all these networks share some principles and technology. First, all implementations assume that every participant of the network can transfer data in any direction. Second, every implementation performs a consistency check of the data by calculating and comparing hash values of transferred pieces of data. These hash values and other knowledge about the data being transferred are called metadata. However, the way this metadata is distributed and discovered varies in different peer network implementations. BitTorrent, the technology we’ve chosen, features several ways to do this. The metadata can be written to a file, called torrent-file, and transferred to the peer by other means, including offline ones. Or, the metadata can be formatted as an URL with a protocol called “magnet” and published on the web or transferred by other means. Above this, there’s ‘broadcatching’. This technology automates metadata distribution and cuts the required time for distribution to a minimum. Also, BitTorrent has sufficient open software support that allows us to use ‘broadcatching’ as a near-real-time data transfer mechanism.
In general, BitTorrent software performs several tasks to enable data sharing. 

First, it prepares the metadata of data to be shared. Besides the name, description, and other fields, the metadata includes a hash value calculated for the whole span of data called “infohash” in BitTorrent and hash values for each piece of the data.

Second, the software maintains a list of peers. This list is populated in several ways: using special software called the Torrent tracker, discovering peers on the local network with broadcast or with multicast, participating in a distributed hash table, and manually. Torrent tracker is a separate piece of software that collects infohashes and peer addresses. When a peer queries the tracker on a particular infohash it responds with a list of peers that might have had the data that suits the provided infohash. Local peer discovery allows the discovery of network stations that run BitTorrent software by broadcasting specific packets on the local network and specific multicast addresses. It is described in BitTorrent specification~\cite{bep14}\cite{bep26}. Distributed Hash Table is a software system that runs on a number of nodes on the network and provides lookup services. Almost any BitTorrent peer participates in DHT and publishes known infohashes to it. Other peers may perform a lookup for the specific infohash and obtain a list of peers, just like tracker software does. Manual mode allows to force the software to connect to the specified address and can be used in conjunction with other peer discovery software or catalog, or for testing.

Third, the BitTorrent software downloads and uploads data. When new an infohash is added to the running instance the software asks known peers whether they have data that suites the infohash. If a peer has this data the software establishes a connection to that peer and downloads the rest of the metadata (piece hashes, filenames, etc.) and the data piecewise, with every piece checked with hash provided in the metadata. When part of the data is downloaded and verified the peer can also serve this part of the data to other peers. These processes are performed simultaneously.

In this paper we utilize broadcatching or “Torrent RSS Feeds”~\cite{bep36}. It is a feature of BitTorrent software that allows the automatic distribution of new content with the help of RSS feeds. RSS (RDF Site Summary or Rich Site Summary) is a structured representation of essential data formatted in XML. This representation is strictly-formatted so that any software that interprets RSS will do so in the same way and will provide similar results. We utilize this mechanism to feed the BitTorrent software of the data consumer with the infohash of the newly available data. According to ~\cite{4784862}, this technology offers fast and low-latency data transfer.
\section{Proof of Concept}

\begin{figure*}
  \includegraphics[width=0.75\textwidth]{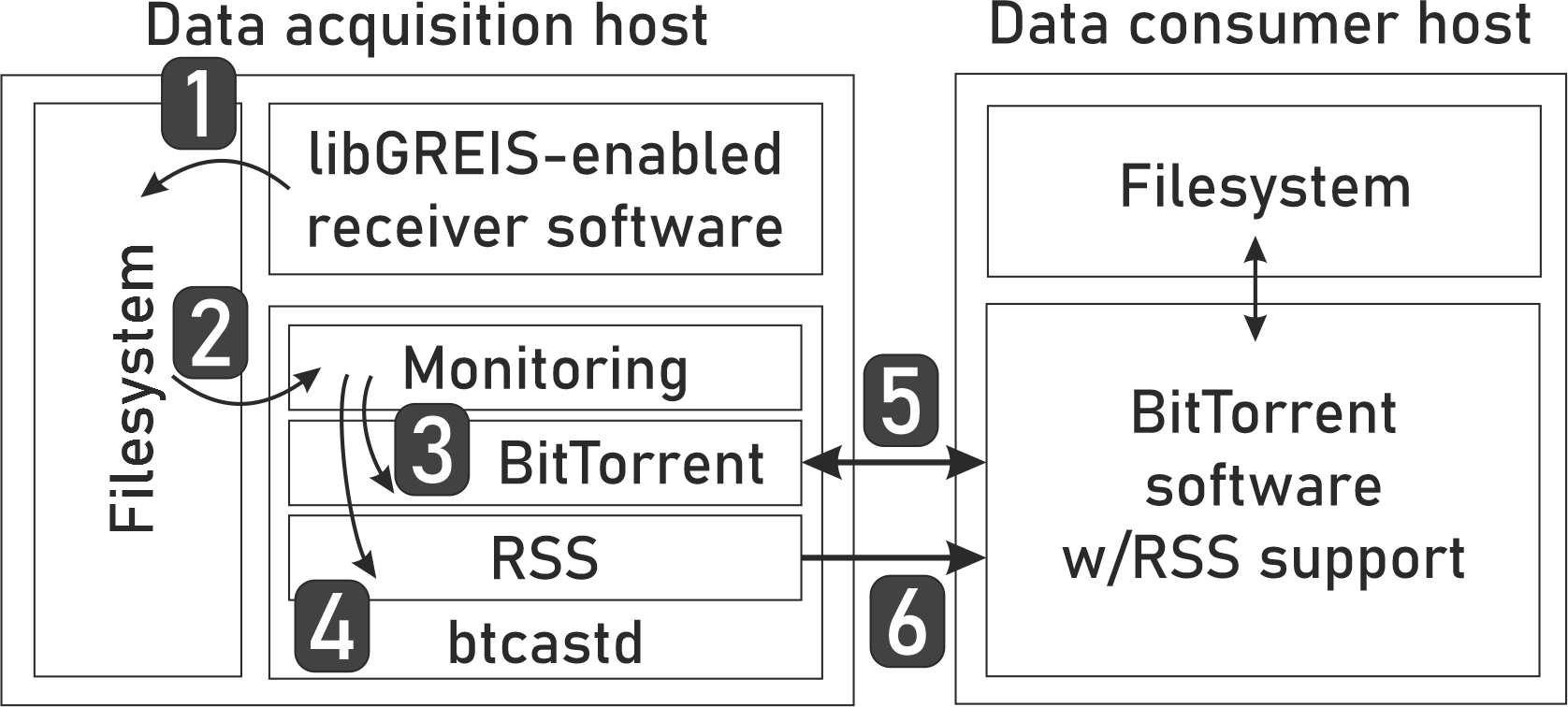}
\caption{The diagram of the software composition on data acquisition host and data consumer host. Data consumer host is generally a typical BitTorrent client, while data acquisition host has special software designed for that role. Running at least two separate processes, the data acquisition software, and btcastd, the whole stack performs the following tasks (numbers correspond to those on the figure): 1. Acquired data is written to the filesystem; 2. The data is read by monitoring part of btcastd; 3. Data is fed to BitTorrent part; 4. Data is fed to RSS part; 5. BitTorrent data transfer operation takes place; 6. RSS is read by BitTorrent software on a data consumer host. The numbers do not depict the order of operation.}
\label{fig:1}       
\end{figure*}

To investigate whether the idea is viable for use, we developed software that implements peer-to-peer data transfer of scientific data. In a word, it catches newly collected data from the GNSS Receiver with the help of~\cite{aleshin2020framework}, computes infohashes, enables data transmission, and presents the RSS feed to potential peers. The software is built upon libtorrent (http://libtorrent.org/), Qt, Mongoose, and is called btcastd (BEE-TEE-cast-DEE). The three major units comprise btcastd: filesystem monitoring and event handling, BitTorrent operations, and on-demand RSS-feed generation and serving, see Fig.~\ref{fig:1}. 

The filesystem monitoring portion detects changes to the specified file directory and adds information on new files to the BitTorrent part when these new files are no longer written to. The latter is essential because the infohash of the file is content-dependent. BitTorrent does not allow the infohash to change over time. This way, only finished non-changing files are allowed into the BitTorrent portion of the software. To achieve this sort of picking mechanism we relied on particular features of the data collection software. Based on libGREIS the GNSS receiver software has a deep view of the data coming in. Knowing the exact time of data chunks it spreads data sequentially into small files. Files are named after the beginning of the N-minute period. Thus, when the current time passes over to the next N-minute file, the current one is no longer written to, hence it could be added to the BitTorrent portion. Here, we reference this method as “time-dependent”. We leverage filesystem change notification features of Qt to learn of new files. There are many other ways of picking files including passage of notifications from data collecting software, monitoring of low-level file handles, and exercising other methods of such monitoring. However, we’ve chosen the time-dependent method for the sake of simplicity. 

The BitTorrent part performs all actions required to make the file available to other BitTorrent peers. It handles infohash computation, negotiations with peers, data transfer, and managing all network operations except for RSS feed. Most operations are handled by libTorrent. 

RSS feed is provided by an integrated HTTP-server powered by Mongoose library. The list of infohashes is updated whenever the infohash is added to the BitTorrent portion of the software. The RSS component just serves a specially crafted RSS feed to potential clients. 

The software was designed to work in conjunction with Greis-enabled GNSS receivers through a data collecting software based on libGreis. Data is collected from receiving hardware and put into files according to a scheme. The scheme allocates file names with the number of minutes since the beginning of the day. These files are put into folders labeled with the day of the year. Next, the days are organized into years. Years comprise the topmost folder in this hierarchy. Btcastd monitors these folders and picks all finished files while maintaining the hierarchy path. To save this hierarchy, the path is recorded into the filename portion of the metadata. The resulting RSS feed can be processed with generic BitTorrent software that is RSS-enabled. During the tests we used qBittorrent. qBittorrent was set up to monitor the address of the host where btcastd had been running. When a new part of data appeared on the RSS feed qBittorrent started to download this file. We relied on local peer discovery for btcastd and qBittorrent to locate each other, although in large-scale deployments use of several torrent trackers is advised. At some point, the data acquisition host has to delete old data to make storage space available for new measurements. When this happens the corresponding torrents are removed from the RSS feed, thus are no longer available from the original location. Not but the relay host and the data center still possess these torrents and associated data. Furthermore, the data center could also create a sort of filing from several pieces of data by generating a new torrent with all those files in it. This will not produce additional copies of data because the hash values of the files still match between the original torrent with only one file and the larger one, but will make it better organized and even more accessible.
\section{Conclusion}

\begin{figure*}
  \includegraphics[width=0.75\textwidth]{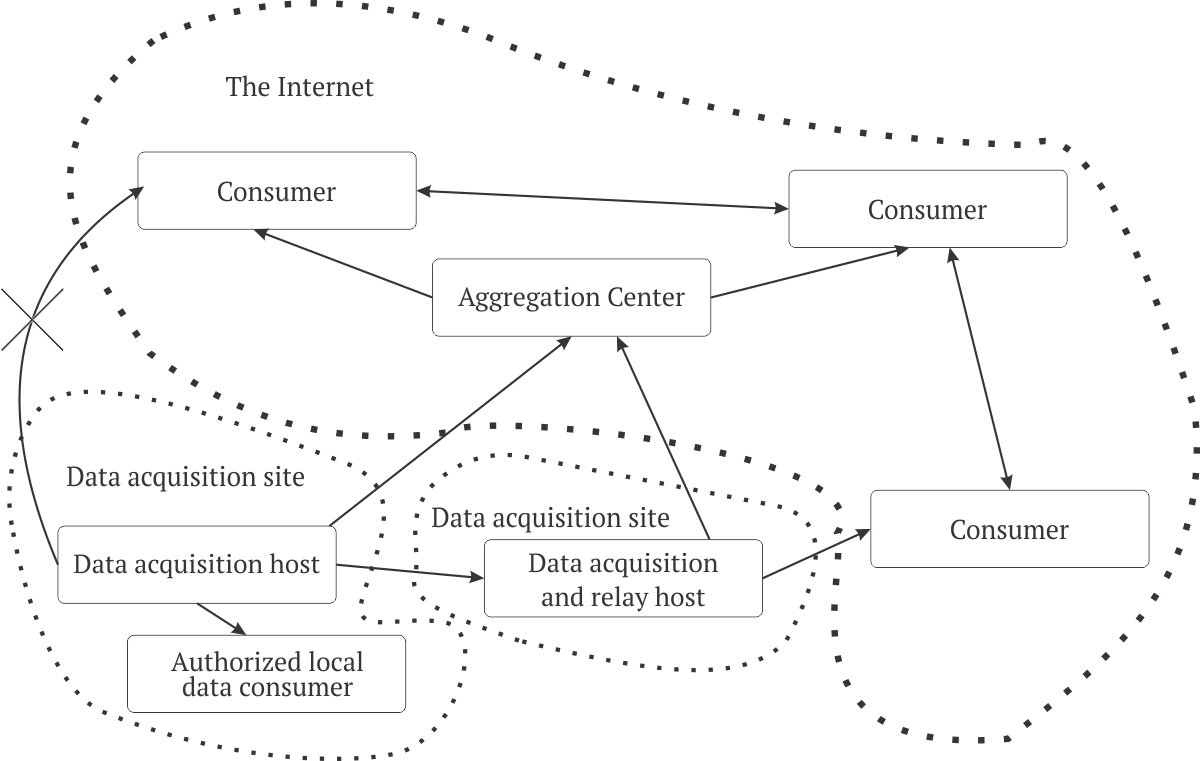}
\caption{Diagram of potential data flows. Data flow is chained from the data acquisition host to the consumers via the data center and optional relays. Note the possible local connection to the authorized local consumer. Blacklisted connections are marked with X. These connections are intentionally disabled to prevent bandwidth abuse at the data acquisition site, although this is optional.}
\label{fig:2}       
\end{figure*}

Presented software could be used to build a sophisticated data handling infrastructure. Within this infrastructure data acquisition hosts (e.g. stations) can reside in remote locations with a suboptimal internet connection and still provide data to the data center or to one of the relay hosts that run BitTorrent software.  To minimize the load on the data acquisition hosts the solution can implement whitelisting of peers on the data acquisition host side that prevents any connections except for relays, data centers, and authorized consumer stations e.g. a local watch office can still source data directly from the data acquisition host. Relays run BitTorrent software and collect data from one or several stations to make it available for the data centers, or if the relay has plenty of bandwidth to serve data to consumers on the internet. Due to the nature of peer-to-peer networks, consumers will get the data at a high speed from several locations including other consumers. The diagram is shown in Fig. ~\ref{fig:2}

This solution has been tested with IPE RAS Data Aggregation Center and experimental GNSS station located in Moscow, Russia. We used libGREIS-enabled GNSS recording software with Javad Alpha 2 GNSS receiver running on a Raspberry Pi 3 board. Acquired files were processed by btcastd and a non-graphical instance of qBittorrent was set at the Center to get all that data with BitTorrent. Consumers were able to collect data similarly. After weeks of operation, the proof-of-concept was deemed viable.

The software implementation is not complex, most of the software could be used out-of-box, e.g. qBittorrent fits well for the consumer role. LibTorrent is simple enough to be used for basic BitTorrent operations, however, things can get quite complicated when it comes to fine-tuning or using advanced features of the library, such as memory and storage I/O management. Diving into these parameters would allow us to combat performance and resource utilization issues to allow smooth operation even on low-power system-on-chip computers although advanced usage is outside the coverage of this article.

It is worth noting that despite the fact that the specification declares the specific way to construct the RSS many BitTorrent implementations construct and parse it in slightly different ways and users may encounter difficulty using broadcatching (e. g. \href{https://github.com/qbittorrent/qBittorrent/issues?q=RSS}{RSS Issues of qBitTorrent on GitHub}) until their or counterpart software is updated with bugfix. However, that’s again a no stopper because the majority of BitTorrent software is open source one can fix or workaround the issue in situ.

We hope to further develop the idea and software implementation to enrich the functionality of applicable data acquisition activities.

See source at \href{https://github.com/iperas/omnicollect}{https://github.com/iperas/omnicollect}

\bibliography{bib}

\end{document}